\documentclass[
superscriptaddress,
 amsmath,amssymb,
 aps,
 pra,
 twocolumn
]{revtex4-2}
\usepackage[utf8]{inputenc}
\usepackage[T1]{fontenc}
\usepackage[autooneside=false]{scrlayer-scrpage}
\usepackage[english]{babel}
\usepackage{amsmath,bm}
\usepackage{amssymb}
\usepackage{siunitx}
\usepackage{listings}
\DeclareSIUnit\eVperc{\eV\per\clight}
\DeclareSIUnit\clight{\text{\ensuremath{c}}}
\usepackage{amsfonts}
\usepackage{graphicx} 
\graphicspath{{./images/}}
\usepackage{hyperref} 
 \hypersetup{
  linktoc = all,
  linkcolor = black,
 }
\usepackage{cleveref}

\usepackage{float}
\usepackage{braket}
\usepackage{nicefrac}
\usepackage{upgreek}
\usepackage[dvipsnames]{xcolor}
\usepackage{tikz,pgfplots}
\usetikzlibrary{positioning}
\usepackage{pgfgantt}
\usetikzlibrary{quantikz}
\usepackage{microtype}
\usepackage{lmodern}
\usepackage{textcomp}
\usepackage{ragged2e}
\usepackage[marginal,norule]{footmisc}
\usepackage{amsthm}
\usepackage{xkvltxp}
\usepackage[draft,inline,margin=false]{fixme}
\fxsetup{theme=color,author=}
\makeatletter
\let\@fnsymbol\@arabic
\makeatother

\usepackage{quoting}
\usepackage{csquotes}

\usepackage{mathrsfs}

\usepackage{dcolumn}
\usepackage{bbold}
\usepackage{color}
\usepackage{booktabs,eqparbox}
\usepackage{pgfplots}
\DeclareUnicodeCharacter{2212}{−}
\usepgfplotslibrary{groupplots,dateplot}
\usetikzlibrary{patterns,shapes.arrows}
\pgfplotsset{compat=newest}
\usepackage{dsfont}
\usepackage{enumitem}

  \usepackage{ulem}

\definecolor{mypurple}{HTML}{CC00CC}
\definecolor{myorange}{HTML}{FFCC99}
\definecolor{mygreen}{HTML}{BFFFBF}
\definecolor{myred}{HTML}{FF0000} 

\begin{document}

\title{Highly Efficient Encoding for Job-Shop Scheduling Problems and its Application on Quantum Computers}

\affiliation{Physics Department, Friedrich-Alexander-Universität Erlangen Nürnberg, Germany}
\affiliation{Siemens Technology, Munich, Germany}

\author{Mathias Schmid}
\affiliation{Physics Department, Friedrich-Alexander-Universität Erlangen Nürnberg, Germany}
\email{mathias.schmid@fau.de}

\author{Sarah Braun}
\affiliation{Siemens Technology, Munich, Germany}
\email{sarah.braun@siemens.com}

\author{Rudolf Sollacher}
\affiliation{Siemens Technology, Munich, Germany}
\email{rudolf.sollacher@siemens.com}

\author{Michael J. Hartmann}
\affiliation{Physics Department, Friedrich-Alexander-Universität Erlangen Nürnberg, Germany}
\email{michael.j.hartmann@fau.de}

\date{\today}

\begin{abstract}
Combinatorial optimization problems are considered to be an application, where quantum computing can have transformative impact. In the industrial context,  job shop scheduling problems that aim at finding the optimal schedule for a set of jobs to be run on a set of machines are of immense interest. Here we introduce an efficient encoding of job shop scheduling problems, which requires much fewer bit-strings for counting all possible schedules than previously employed encodings. For problems consisting of $N$ jobs with $N$ operations, the number of required bit-strings is at least reduced by a factor $N / \log_2(N)$ as compared to time indexed encodings. This is particularly beneficial for solving job shop scheduling problems on quantum computers, since much fewer qubits are needed to represent the problem. Our approach applies to the large class of flexible and usual job-shop scheduling problems, where operations can possibly be executed on multiple machines. Using variational quantum algorithms, we show that the encoding we introduce leads to significantly better performance of quantum algorithms than previously considered strategies. Importantly, the encoding we develop also enables significantly more compact classical representations and will therefore be highly useful even beyond applicability on quantum hardware. 
\end{abstract}

\maketitle

\section{Introduction}

Combinatorial optimization problems are regarded as a promising use case for quantum computing. The intuition behind most considered approaches, such as quantum annealing \cite{Kadowaki98} and the Quantum Approximate Optimization Algorithm (QAOA) \cite{QAOA} is that quantum computers can examine many configurations of the considered variables in parallel and can therefore provide a significant advantage. First demonstrations of quantum algorithms for such tasks have recently been reported for various quantum hardware platforms \cite{harrigan2020,Pagano2020,Lacroix2020,Graham:2022yu,Ebadi2022}.

A combinatorial optimization problem, which is of paramount relevance in industry, computation, and economics, is job shop scheduling \cite{fjsp_review_2014,fjsp_review_2018}. It considers a set of jobs of varying processing times and seeks to schedule them on a set of machines and/or operators with varying processing capabilities such that the makespan, the time it takes to complete the last finishing job, is minimized. The job shop scheduling problem (JSP) occurs in various contexts, ranging from finding the optimal schedule for a set of production jobs in a given industrial facility to train scheduling optimization \cite{train_scheduling_review} and optimizing the scheduling of compute jobs in a computing cluster. By solving the corresponding optimization problem in all these applications, idle times of machines or computers can be reduced, the economic and energetic efficiency of a production or computing cluster can be increased, and transportation systems can operate more reliably at even higher throughput.

Due to its extraordinary relevance, significant effort has been put into solving the JSP \cite{FJSPreview2014,fjsp_review_2018}. Approaches on classical computers usually require a large number of variables. In quantum computing, this is however not practical, and reducing the number of variables to represent the problem becomes crucial for solving JSPs.

There are already quantum approaches to the JSP, which mainly consider casting the problem into a quadratic unconstrained binary optimization (QUBO) formulation \cite{jsp_qc_review}. The most often considered formulation is the time indexed version \cite{Venturelli, filtering_case_study, timeindex2,timeindex3, timeindex4, timeindex5, timeindex6,timeindex_qaoa}, where one assigns a set of binary variables to each
operation, corresponding to its various possible
starting times. This encoding provides the benefit that it can be used with a quantum annealer\cite{Venturelli, timeindex2, timeindex4,timeindex5, timeindex6}, but on the other hand, the number of possible starting times, i.e the variable count, scales rather badly. Consequently, a related approach for a digital quantum computer, using QAOA, struggles already for toy instances with only six operations and small execution times \cite{timeindex_qaoa}. This problem of the large number of required variables becomes even more pronounced when generalizing the problem to the flexible job shop problem (FJSP)\cite{timeindex3, timeindex5}, which is a generalization of the JSP, where the operations within each job can possibly be executed on various machines.

Here we will forgo a QUBO formulation of the JSP or FJSP and instead use a black box approach, where the cost function, in our case the makespan, is not represented via a quantum observable but can be calculated from a measured output of the quantum processor in polynomial time. Such unconstrained black box binary optimization was recently also investigated for feature selection \cite{blackbox_hamiltonian_feature}. 

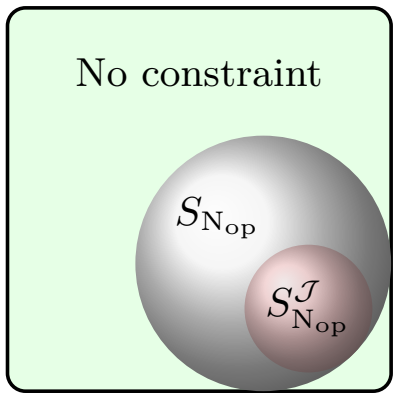
\begin{figure}[ht!]
    \centering
    \scalebox{1.7}{
\begin{tikzpicture}[]
  \filldraw[draw=black,thick,fill=green!10!, rounded corners] (-2, 2) rectangle (1,-1);
  \draw node at (-0.5, 1.5){No constraint};
  \shade[ball color=gray!10!] (0,0) circle (1);
  \draw node at (- 0.3535, 0.3535){$S_\mathrm{N_{op}}$};
  \shade[ball color=red!20!, fill opacity=0.75, text opacity= 1] (0.3535,-0.3535) circle (0.5) node{$S^\mathcal{J}_\mathrm{N_{op}}$};
\end{tikzpicture}}
    \caption{Pictorial representation of the different code spaces. Respecting the constraints of the problem in its encoding reduces the code space. $N_\mathrm{op}$ is the number of operations in a given JSP instance, $S_{N_\mathrm{op}}$ the set of permutations or symmetric group on $N_\mathrm{op}$ elements, and $S_{N_\mathrm{op}}^\mathcal{J}$ the symmetric group without permutations that violate the job precedence constraint. See \cref{sec:problem_def} for the detailed discussion.}
    \label{fig:encoding_sets}
\end{figure}

For JSPs and FJSPs, this choice of cost function allows us to derive encodings that respect the constraints given by the problem within the encoding. The considered problems can therefore be represented on quantum computers with much fewer qubits than in previously considered approaches, thus significantly lowering the requirements for solving JSPs and FJSP on quantum computers. The objective of our encoding is illustrated in \cref{fig:encoding_sets}. 

Using this encoding in variational quantum algorithms of course requires the classical routine used to optimize a parameterized quantum gate sequence to be compatible with such a choice of cost function. We here use the recently proposed filtering variational quantum eigensolver (F-VQE) \cite{filtering}, as it does not require formulating the cost function of the problem as a Hamiltonian. In combination with our efficient encoding strategy, this approach allows us to test the quantum algorithm in numerical simulations for much larger instances than previously considered, showing that our new problem representation leads to significantly improved performance of near-term quantum algorithms. We however stress that our main result, the new efficient encoding we developed, can be employed in quantum and classical algorithms so that its benefits are largely independent of the progress in quantum hardware development.

In the following we first define the class of problems that we consider before introducing the encoding we developed to represent these problems. We then conclude with numerical experiments showing the efficiency of the resulting quantum algorithms.

\section{Problem definition}\label{sec:problem_def}
A job shop scheduling problem consists of a set of $N_{\text{jobs}}$ jobs \(\mathcal{J}=
  \{J_1,\cdots,J_{N_{\text{jobs}}}\}\) which are to be scheduled on a set of $N_\text{ma}$ machines \(\mathcal{M}=\{M_1,\cdots,M_{N_\text{ma}}\}\).   Each job $J_i$ consists of a sequence of $n_i$ operations that must be executed in predefined order,
\begin{align} \label{eq:lexicographical}
   J_i&=\{O_{k_{i-1}+1}\rightarrow\cdots\rightarrow O_{k_i}\} \quad \text{for} \quad i = 1, \dots, N_{\text{jobs}}
\end{align}
where we use an enumeration of the operations in lexicographical order, c.f. \cite{Venturelli}, and $n_i = k_i - k_{i-1}$. The predefined order in \cref{eq:lexicographical} is called the job precedence constraint.
Moreover, each operation \(O_{i}\) has to be executed by one machine from an operation specific set,
  \begin{equation*}
  \mu_{i}=\{M_{i,1},M_{i,2},\cdots\};  \quad M_{i,j}\in\mathcal{M},
\end{equation*}
and has an execution time \(p_{i,l}\), that depends on the machine $M_l$ it is executed on.
We denote the set of execution times by \(\mathcal{P}\).
There are two versions of the problem. In the standard job shop scheduling problem (JSP), there is only one machine that can execute an operation, $|\mu_i| = 1$. In contrast, in the flexible job shop scheduling problem (FJSP), there are operations that can be executed on multiple machines, $|\mu_i| \geq 1$.

The objective of the problem is to find a schedule of operations and a choice for the machines such that the finishing time of the last job is minimal. This time is called the makespan $C_\mathrm{max}$. We denote the finishing times of the individual jobs $J_i$ by $C_i$ such that $C_\mathrm{max}=\max_{i\in\{1,\cdots,N_{\text{jobs}}\}} C_i$.
We note that there are alternative objective functions for the FJSP which we will not elaborate upon \cite{FJSPreview2014,fjsp_review_2018}.
We make the following, rather natural assumptions:
\begin{enumerate}
    \item All machines are available at time zero and all jobs can start at time zero.
    \item The order of operations in each job is predefined and cannot change, see \cref{eq:lexicographical}.
    \item Operations of different jobs have no precedence constraint among each other and can also be processed in parallel.
    \item The execution times for all operations on each machine are defined in advance.
    \item A job can use a machine more than once.
    \item Each machine can only work on one operation at a time. This is the machine constraint
    \item A machine cannot be interrupted during the execution of a operation.
\end{enumerate}
After having introduced the job shop problems we consider, we now turn to discussing how to efficiently encode them in binary numbers that can be represented by qubits on a quantum computer.

\section{Encoding}\label{sec:encoding}
To represent a combinatorial optimization problem on a quantum computer, one typically needs one qubit for each variable of the problem. Hence formulating the problem with as few variables as possible is very beneficial. Here we introduce an encoding for FJSPs (and JSPs), that is much more variable efficient than previously considered encodings \cite{Venturelli, filtering_case_study, timeindex2,timeindex3, timeindex4, timeindex5, timeindex6,timeindex_qaoa}. 

The encoding is based on a operation sequencing list \(S\) \cite{sequenced_encoding,sequenced_encoding_origin}, which orders all operations together with the machines they are executed on,
\begin{widetext}
\begin{equation} \label{eq:schedule}
    S= (O_{\pi(1)},M_{\pi(1),j_1}), (O_{\pi(2)},M_{\pi(2),j_2}), 
    \dots, (O_{\pi(N_{\text{op}})},M_{\pi(N_{\text{op}}),j_{N_{\text{op}}}}), 
\end{equation}
\end{widetext}
where $N_{\text{op}}$ is the total number of operations, $\pi$ a permutation of the operations that leads to the schedule and $M_{\pi(i),j_i} \in \mu_{\pi(i)}$.
The notation used in \cref{eq:schedule} is to be understood such that starting from the left, each operation in $S$ is executed on a corresponding machine as early as possible, i.e. when the machine becomes available or the preceding operation in the same job has been completed. 
Before discussing the representation of sequencing lists in terms of bitstrings or qubits we first show that they represent valid schedules and allow for an efficient computation of their makespan.

 \subsection{Cost function}\label{subsec:cost}
The cost function we want to optimize is the makespan, which is equal to the finishing time of the last active job (The latter is not necessarily the last job in the sequence \cref{eq:schedule}).
To evaluate the makespan of a given schedule we will use an algorithm, which iteratively determines the starting times of each operation within the schedule before adding the respective execution times and taking the maximum.

\begin{figure*}[t]
\begin{minipage}[c]{0.6\linewidth}
\begin{itemize}[leftmargin=0em]
    \item[\textbf{a)}] 
    \hphantom{a}\\
\includegraphics[width=1\textwidth]{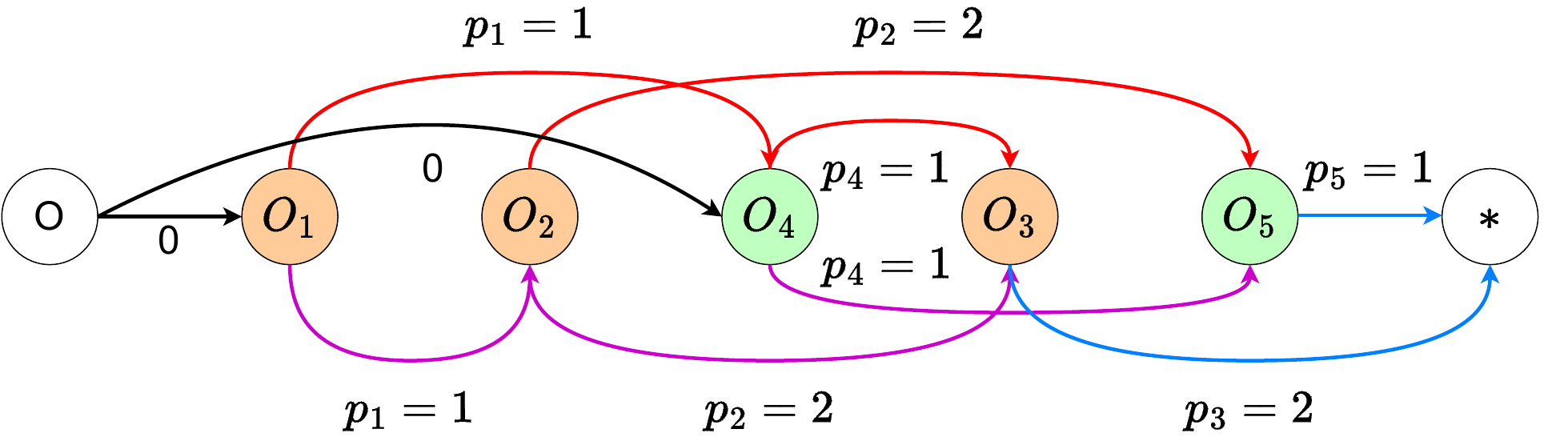}
\vspace{0.08cm}
\end{itemize}
\end{minipage}
\begin{minipage}[c]{0.35\linewidth}
\begin{itemize}[leftmargin=0em]
\item[\textbf{b)}]
\hphantom{a}\\
\vspace{-0.4cm}
\input{images/gantt}
\end{itemize}
\end{minipage}
\caption{\textbf{a)} An example schedule $S=(O_1,M_1),(O_2,M_2),(O_4,M_1),(O_3,M_1),(O_5,M_2)$ depicted as graph for an example instance with five operations in two jobs, $J_1 =\{O_1,O_2,O_3\}$ and $J_2=\{O_4,O_5\}$, two machines \(\mathcal{M}=\{M_1,M_2\}\), and the execution times $p_j = 1,2,2,1,1$ for each operation $O_j$ $(j=1, \dots, 5)$. The starting time of operation $O_j$ is the longest path from the origin to operation $O_j$ and the makespan is the longest path from the origin to the sink. Operations in orange are part of job $J_1$, while green operations are elements of job $J_2$. Purple edges represent a delay caused by the job precedence constraint, c.f. \cref{eq:lexicographical}, red edges represent a delay caused by the occupation of the corresponding machine, black edges are connecting the origin to the first operations within each job and blue edges represent the execution times of operations which are last in their job as well as the last operations processed on their machine.
Following the algorithm described by \cref{eq:starting-times}, one here observes that 
the schedule starts with $O_1$ at \(t_1(1)=0\). Then, the operation in sequence position two ( $s=2$, here \(O_2\)), is delayed by the first operation due to the job constraint, i.e. one has to wait until the preceding operation in the job has finished. The starting time is \(t_2(2)=1\). The third place operation ($s=3$, here \(O_4\)) is delayed by the machine constraint since $M_1$ was in use and the starting time becomes $t_3(4)=1$. Similarly we obtain $t_4(3)=3$ and $t_5(5)=3$. Adding the execution times and taking the maximum leads to a makespan of 5. With this scheme one can also construct a Gantt chart, in
\textbf{b)},  of the example schedule. Time runs from left to right (columns) and the machine load is depicted along the rows.}
\label{fig:path2_2}
\end{figure*}

Starting with $i=0$, operation $O_j$ in schedule position $s(j) = i$ has a starting time $t_i(j)$ given by
\begin{equation} \label{eq:starting-times}
    t_i(j) = \max (t_f(j-1), t_{fm})
\end{equation}
where $t_f(j-1)$ is the time at which operation $O_{j-1}$ finishes if it is in the same job (otherwise we choose $t_f(j-1)=0$). The time $t_{fm}$ is the finishing time of the last operation, scheduled before $O_j$, running on the same machine as $O_j$. If $O_j$ is the first operation running on the machine, $t_{fm}$ is set to $0$.

Thus, we use \cref{eq:starting-times} to iteratively determine all $t_i(j)$. 
The procedure is illustrated for an example in \cref{fig:path2_2}.

Once we have determined the starting times we add their execution times and take the maximum, which results in the makespan, 
\begin{equation} \label{eq:makespan}
    C_{\text{max}} = \text{max}_{j}[t_i(j) + p_j]
\end{equation}
Since the above procedure determines all the starting times of the operations, it can also be used to produce the gantt chart of the schedule, see panel b) in \cref{fig:path2_2} for the considered example.

After confirming that a sequencing list represents a valid schedule and allows for an efficient computation of the makespan, we now turn to explaining our main result and show how sequencing lists translate efficiently onto bitstrings and thus qubits.

\subsection{Mapping to bitstrings (qubits)}
The representation of the schedule as in \cref{eq:schedule} already respects many constraints given by the problem, e.g. operations of the same job are not scheduled in parallel and the machine constraint is automatically fulfilled. Furthermore, it does not require any further encoding specific constraints.  In contrast, encodings which use cost functions that can be written in QUBO formulation \cite{Lucas_2014} and therefore as a Hamiltonian, e.g. as required in the quantum approximate optimization algorithm (QAOA) \cite{QAOA}, need to add penalty terms to the cost function in order to ensure that job precedence \cref{eq:lexicographical} and machine constraint are fulfilled.

Schedules as in \cref{eq:schedule} moreover do not incur unnecessary machine idle since every operation is started as soon as possible. Therefore, the encoding is for almost all cases, less variable intensive  than the widely used time index version \cite{Venturelli, filtering_case_study, timeindex2,timeindex3, timeindex4, timeindex5, timeindex6,timeindex_qaoa}. Yet, sequencing lists, as in \cref{eq:schedule}, still contain invalid schedules as the job precedence constraint \cref{eq:lexicographical} may be violated by some permutations. 

Here we will describe how to avoid encoding such invalid schedules and efficiently encode the sequenced based schedule such that the job precedence constraint in \cref{eq:lexicographical} is resolved automatically.

Since allowing only one ordering within each job is equivalent to considering all orderings within the job as indistinguishable and counting them as only one possibility, the number of schedules that respect the job precedence constraint is equal to the number of permutations of $N_\mathrm{op}$ operations, where all operations within each job are treated as if they were identical. The number of valid schedules for the JSP, where each operation can only run on one machine, is thus given by the multinomial coefficient 
\begin{equation*}
    \binom{N_\mathrm{op}}{|J_1|,\dots,|J_{N_{\text{jobs}}}|} = \frac{N_\mathrm{op}!}{\prod_k |J_k|!},
\end{equation*}
For the FJSP, one can then obtain the total number of valid schedules by multiplying with the number of possible machines for each operation, $\prod_i |\mu_i|$.
As a result, the encoding we introduce only requires the following qubit numbers,
\begin{align}
    N_\mathrm{qubits} & =\lceil\log_2(\frac{N_\mathrm{op}!}{\prod_k |J_k|!})\rceil \qquad \qquad \, \text{for the JSP} \label{eq:nqubits_JSP}\\
        N_\mathrm{qubits} & =\lceil\log_2(\frac{N_\mathrm{op}!}{\prod_k |J_k|!}\prod_i |\mu_i|)\rceil \quad \text{for the FJSP}, \label{eq:nqubits_FJSP}
\end{align}
which are much smaller than those required for other encodings as we discuss further below. 

For mapping schedules, as given in \cref{eq:schedule}, which respect the job constraint \cref{eq:lexicographical}, to bit-strings, we need a computationally efficient strategy. The algorithm we have developed for this purpose maps the schedules, which are permutations, to their inversion vectors, which allow to identify and discard invalid schedules in a straight forward way. For enumerating the remaining valid schedules, we have designed a mixed radix number system, inspired by the factorial number system, which can be converted to binaries using standard procedures. The details of this algorithm are presented in the methods \cref{sec:methods}.

One could reduce the number of required variables further by breaking inherent symmetries.  Two adjacent operations in $S$, that are in different jobs and are executed on different machines are interchangeable \footnote{This means 

$\cdots,  (O_{\pi(i)},M_{\pi(i),j_i}), (O_{\pi(i+1)},M_{\pi(i+1),j_{i+1}}) \cdots 
\equiv \cdots(O_{\pi^{\prime}(i)},M_{\pi^{\prime}(i),j_i}), (O_{\pi^{\prime}(i+1)},M_{\pi^{\prime}(i+1),j_{i+1}})\cdots$, 
where $\pi(i)=\pi^{\prime}(i+1)$, $\pi(i+1)=\pi^{\prime}(i)$, $M_{\pi(i),j_i}\neq M_{\pi(i+1),j_{i+1}}$, $O_{\pi(i)}\in J$ but $O_{\pi(i+1)} \notin J$ and $\pi(k)=\pi^{\prime}(k)$ for $k\neq i ,i+1$.}.
The encoding we introduce, is optimal in terms of variable efficiency up to these possible symmetry reductions. Moreover, it directly resolves all problem constraints and thus avoids the need for any penalty terms in the cost function to enforce these. Due to these properties, our encoding provides a significant improvement in terms of variable efficiency.

For example, the time index encoding of the JSP, without variable pruning, requires $N_\mathrm{op}T$ binary variables, where $T$ is the number of discrete time steps at which an operation can start. $T$ can be lower bounded by the largest sum of execution times for an individual job, which itself is a lower bound for the makespan. After variable pruning the required number of qubits is
\begin{equation}N_\mathrm{qubits}^\mathrm{ind}=N_\mathrm{op}T-\sum_i\sum_{l}(p_l|J_i|-1)
\end{equation}
with $i\in\{1,\dots,N_\mathrm{jobs}\}$ and $l\in\{k_{i-1}+1,\dots,k_i\}$.
In contrast, the required number of qubits in our encoding is independent of the resolution used to measure the durations of the individual operations. Thus when considering the common example of a JSP with $N$ jobs, where each job consists of $N$ operations, i.e. $N_\mathrm{op}=N^2$, the encoding strategy we introduce requires $N_\mathrm{qubits}=\lceil \log_2(N^2!/{(N!^N}))\rceil \approx N^2 \log_2 N$ qubits, whereas the number of qubits for the time indexed version is proportional to $<p>N^3 + N^2$, where $<p>$ is the average of all $p_i$. Hence our encoding requires a factor of  $\sim N / \log_2(N)$ less qubits. When considering a fixed number of jobs, $N_\mathrm{jobs}$, where the system size grows only with job size $|J_i|=N$, our qubit count scales linearly with $N$ compared to $N^2$ for the time indexed version. 
For total flexible job shop problems, where $\mu_i=\mathcal{M} \quad \forall i$ and the number of machines grows linearly with system size, the advantage becomes even larger.
For both cases, $N$ jobs with $N$ operations each or a fixed number of jobs, the reduction becomes $\sim N^2/{\log_2(N)}$. A more detailed discussion of the qubit count behaviors for various problem configurations can be found in \cref{sec:appendic_scaling}.
    
There are also other possible direct encodings for a quantum computer, e.g. the disjunctive encoding in QUBO form as described in \cite{disjunctive}. However, since these use multiple integer or discrete variables, a translation onto qubits via binary expansion or one-hot encoding is costly. 

We note that the encoding we introduce could also be used to represent possible travel routes of the traveling salesman problem as permutations. For this representation it would only require $\lceil\log_2(n!)\rceil \sim n \log_2(n)-n$ qubits for routes through $n$ cities, instead of the $n^2$ qubits needed in existing representations \cite{goswami2004efficiently}. Here our construction can take contraints, such as required precedence for the visits to a subset of the cities, into account directly in the encoding and thus leads to a maximally compact representation.

\section{Application in a Quantum Algorithm} \label{sec:simulation}

To illustrate that our approach can successfully be employed to solve JSPs and FJSPs, we use our encoding strategy in a near term quantum optimization algorithm. For this purpose we have chosen the filtering variational quantum eigensolver (F-VQE), see \cref{sec:fvqe} \cite{filtering}, as it does not require formulating the cost function of the problem as a Hamiltonian. 

The F-VQE is a variational quantum algorithm, that optimizes a parameterized gate sequence by constructing operators which filter out the components of a trial state, which have lower cost function values. This filtering operation can thereby be determined for measured bit-strings, see \cref{sec:fvqe} for details, which makes F-VQE a suitable optimizer for our efficient problem encoding.

We generated random instances of both, the JSP and FJSP, see \cref{sec:rand_inst} for details, for which we computed the approximation ratio
\begin{equation} \label{eq:appr-ratio}
    \alpha = \frac{C_\mathrm{gs}}{C_\mathrm{opt}} .
\end{equation}
Here, $C_\mathrm{gs}$ is the minimal achievable makespan, which we find by evaluating all schedules, and $C_\mathrm{opt}$ is the makespan achieved by our algorithm.
To gain further insight, we also computed the probability to measure the ground state,
\begin{equation}\label{eq:ground_state}
    \sum_i |\braket{\psi_\mathrm{opt}|\psi_{\mathrm{gs},i}}|^2
\end{equation}
where $\ket{\psi_{\mathrm{gs},i}}$ is a computational basis state with corresponding makespan $C_\mathrm{gs}$ and $\ket{\psi_\mathrm{opt}}$ is the wave function found by the optimizer.

\begin{figure*}
 \Centering
\includegraphics[width=\textwidth]
{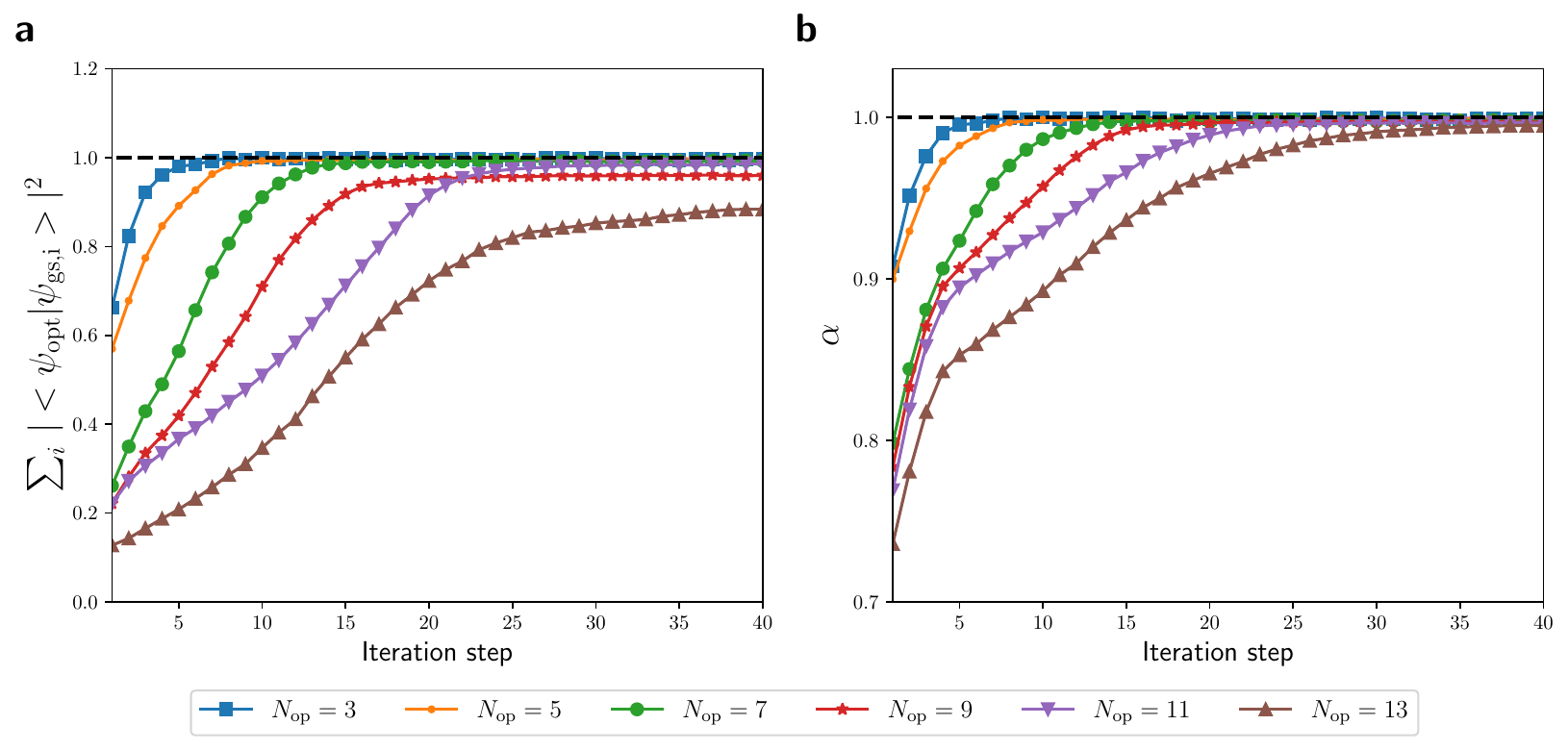}%
\caption{(\textbf{a}) Probability to measure the ground state, see \cref{eq:ground_state} and (\textbf{b}) approximation ratio $\alpha$, see \cref{eq:appr-ratio}, for JSPs with varying number of operations $N_\mathrm{op}$. We consider $N_\mathrm{meas}$ measurements for each set of variational parameters, where for $N_\mathrm{op}<6$ we use $N_\mathrm{meas}=100$, for  $6\leq N_\mathrm{op}<9$ we use $N_\mathrm{meas}=500$ and for $N_\mathrm{op}\geq 9$ we use $N_\mathrm{meas}=1000$. The legend applies to both plots. The data points are averages over 30 random JSP instances, c.f. \cref{sec:rand_inst}.}
\label{fig:permj}   
\end{figure*}

The algorithm converges to the ground state for the vast majority of considered JSP instances, see \cref{fig:permj}. For larger system sizes there were however a few instances, for which the algorithm instead converged to a low lying excited state that still corresponds to a good solution, i.e. an efficient schedule, although not the absolute optimum. We attribute these observations to properties of the FVQE algorithm. Yet, we see several ways for improving the convergence properties. These include (i) taking a smaller learning rate $\eta$, (ii) doing a more exhaustive optimization of the cost function in \cref{eq:cost_t} or (iii) use different filtering operators.

\begin{figure*}
 \Centering
\includegraphics[width=\textwidth]
{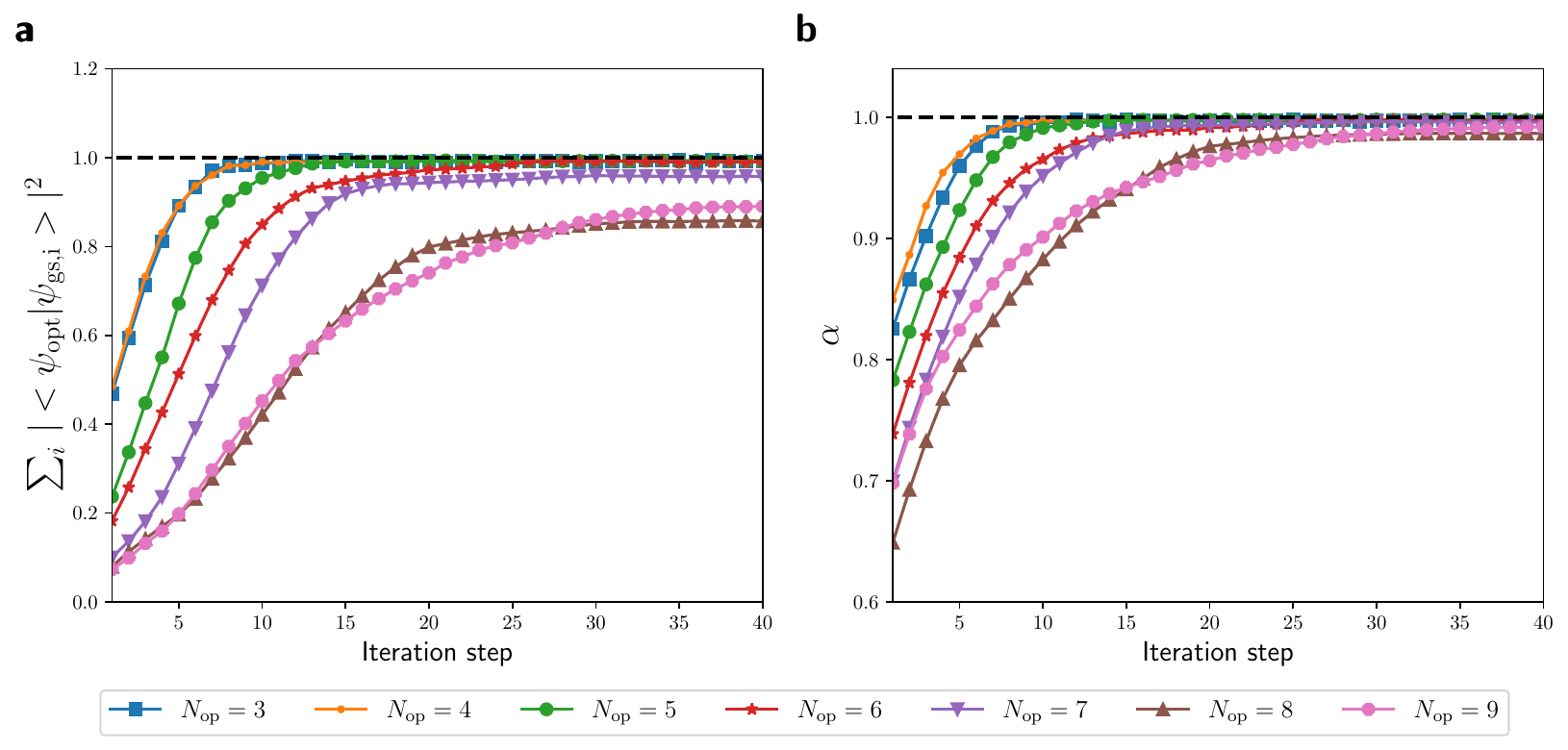}
\caption{(\textbf{a}) Probability to measure the ground state, see \cref{eq:ground_state} and (\textbf{b}) approximation ratio $\alpha$, see \cref{eq:appr-ratio}, for FJSPs with varying number of operations $N_\mathrm{op}$. We consider $N_\mathrm{meas}$ measurements for each set of variational parameters, where  for $N_\mathrm{op}<5$ we use $N_\mathrm{meas}=100$, for  $5\leq N_\mathrm{op}<7$ we use $N_\mathrm{meas}=500$ and for $N_\mathrm{op}\geq 7$ we use $N_\mathrm{meas}=1000$. The legend applies to both plots. The data points are averages over 30 random FJSP instances, c.f. \cref{sec:rand_inst}.}
\label{fig:permj_multm}   
\end{figure*}

For FJSPs, as for the JSPs, we find that the algorithm converges to the ground state for the vast majority of considered instances, see \cref{fig:permj_multm}, with a few exceptions where the algorithm instead converged to a low lying excited state, as for the JSP. We again attribute these to properties of the FVQE algorithm and expect that the same suggestion to improve the performance also apply here. 

\begin{figure*}[!ht]
\Centering
\includegraphics[width=\textwidth]{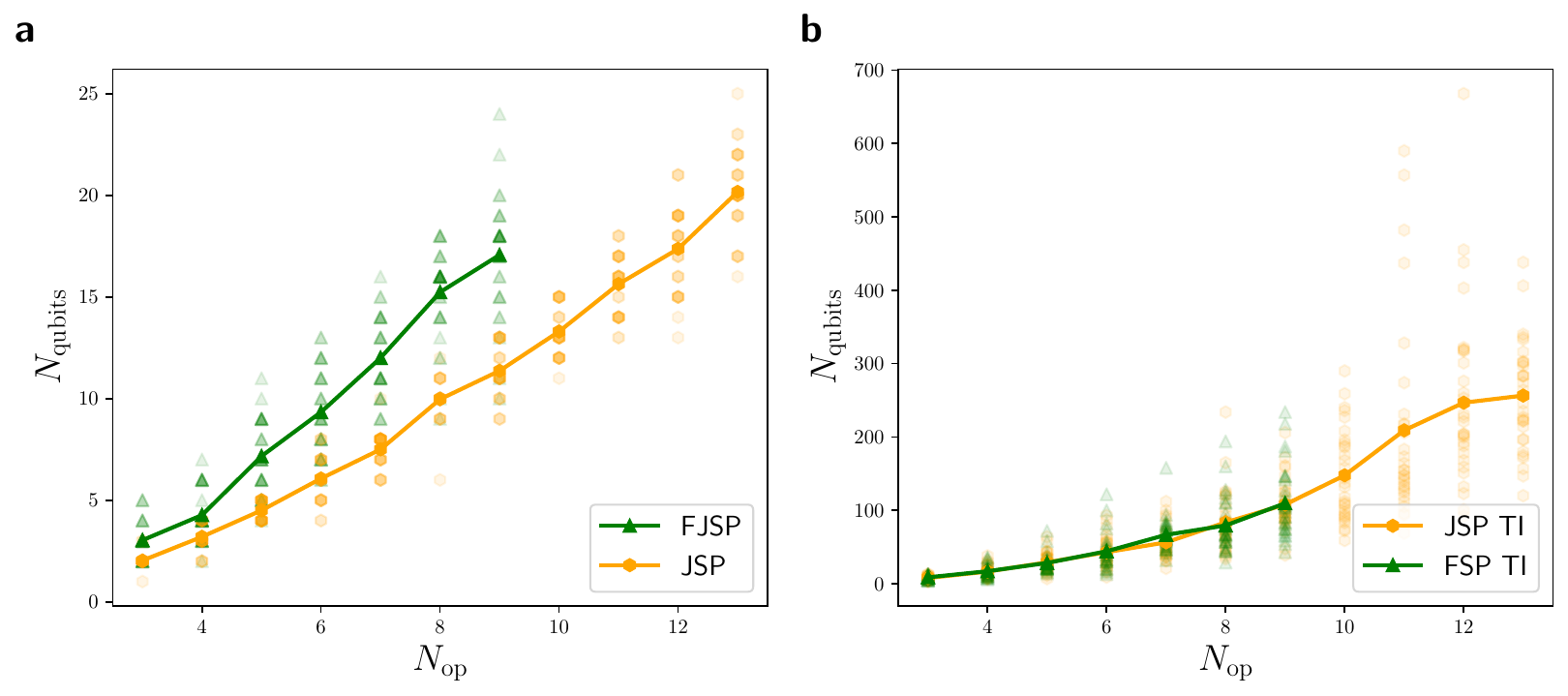}%
\caption{Required number of qubits versus number of operations for our approach  in panel (\textbf{a}) and for the time indexed version with simple variable pruning in panel (\textbf{b}). Note that the ranges for the vertical axes differ by more than a factor 20. The data points connected via lines are averages over 30 JSP (FJSP) random instances, c.f. \cref{sec:rand_inst}. Transparent data points represent individual instances, where the opacity reflects the number of instances that require the same number of qubits (higher opacity meaning more instances). The considered timespan $T$ in (\textbf{b}) is set to its minimal possible value $T = C_\mathrm{gs}$, which leads to the minimal qubit count for this specific time indexed encoding. }
\label{fig:permj_multm_qubits}
\end{figure*}

We also analyzed the number of qubits needed to represent the random problem instances we considered. Fig. \ref{fig:permj_multm_qubits} shows the required qubit numbers versus the number of operations for our approach (panel \textbf{a}) and compares them to the required qubit numbers for the time indexed version with simple variable pruning (panel \textbf{b}), c.f. \cref{sec:appendic_scaling}. For the considered problem sizes, which are rather small so that the quantum algorithm can be numerically emulated classically, we already observe a significant reduction by roughly a factor 10. The FJSP instances require more qubits since it is a more complex problem than the JSP. In the time indexed version this effect is mitigated by the fact that the makespan is reduced through the flexible nature of the FJSP instances.

\section{Conclusions and Outlook}

In summary, we have introduced a novel encoding that allows to represent job shop scheduling problems on quantum computers with much viewer qubits than previously required. This is achieved by identifying schedules with permutations and translating the precedence constraint within jobs to inequalities for the inversion vector elements. Via numerical experiments, we have furthermore shown that this encoding, combined with the hardware and problem agnostic F-VQE  \cite{filtering}, leads to much more efficient quantum solutions than more traditional versions. While such algorithms are a prime candidate in the current NISQ era, we expect our encoding to be equally valuable in fault tolerant quantum computing where the number of logical qubits will also be rather restricted. One possible direction to take in further research is to investigate a method to reduce the inherent symmetries of the encoding and thus further reduce the number of required qubits. It might also be interesting to expand the numerics and focus more on the optimizer performance. There one could also compare different problem agnostic optimization algorithms to find the most suitable. The encoding presented was done for the flexible job shop problem, but the method used can be broken down to a map between a binary and a permutation with or without constraints on inversions. Thus, it is likely to find other problems where this scheme is applicable.

\section{Acknowledgements}
MS and MJH received support from Siemens Technology. This work is part of the Munich Quantum Valley, which is supported by the Bavarian state government with funds from the Hightech Agenda Bayern Plus. RS and SB received support by the German Federal Ministry for Economic Affairs and Climate Action (BMWK) via the funded research project QCHALLenge (01MQ22008B). The authors are seeking patent protection for ideas in this work.


\clearpage
\section{Methods}\label{sec:methods}
In this section we discuss the details of the bijective mapping between bit-strings and valid schedules that we developed. We first encode the order of operations under the assumption of $|\mu_i|=1 \,\forall i$, corresponding to the JSP case, into a bit-string before resolving the job constraint of \cref{eq:lexicographical} in \cref{sec:job_con} and generalizing to the case with $|\mu_i|\neq1 \,\forall i$ in \cref{sec:multm}.

\subsection{Labeling permutations with bit-strings} \label{sec:methods-A}
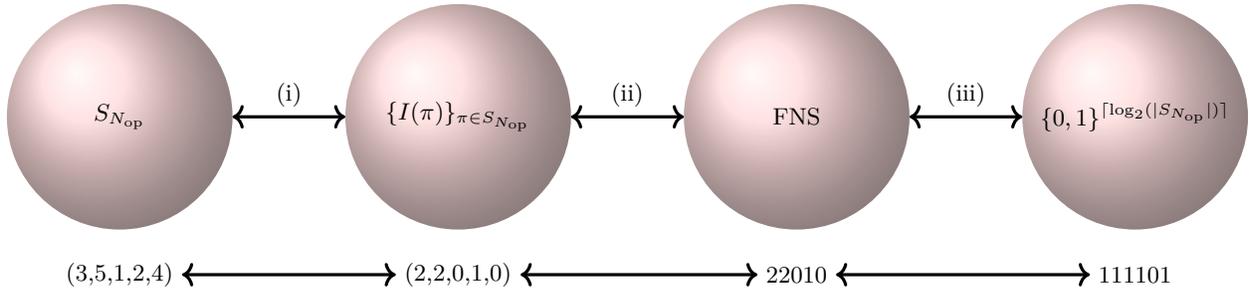
\begin{figure*}[!htbp]
    \centering
    \scalebox{1}{
\begin{tikzpicture}[scale =1.5]
\shade[ball color=red!20!, fill opacity=0.75, text opacity= 1] (0,0) circle (1) node{$S_{N_\mathrm{op}}$};
\draw (0,-1.4) node (A) {(3,5,1,2,4)};
\shade[ball color=red!20!, fill opacity=0.75, text opacity= 1] (3,0) circle (1) node{$\{I(\pi)\}_{\pi\in S_{N_\mathrm{op}}}$};
\draw (3,-1.4) node (B) {(2,2,0,1,0)};
\shade[ball color=red!20!, fill opacity=0.75, text opacity= 1] (6,0) circle (1) node{FNS};
\draw (6,-1.4) node (C) {22010};
\shade[ball color=red!20!, fill opacity=0.75, text opacity= 1] (9,0) circle (1) node{$\{0,1\}^{\lceil\log_2(|S_{N_\mathrm{op}}|)\rceil}$};
\draw (9,-1.4) node (D) {111101};
\draw[<->, very thick](1,0) -- (2,0) node [above,midway] {(i)};
\draw[<->, very thick](4,0) -- (5,0) node [above,midway] {(ii)};
\draw[<->, very thick](7,0) -- (8,0) node [above,midway] {(iii)};
\draw [<->, very thick] (A) edge (B) (B) edge (C) (C) edge (D);
\end{tikzpicture}}
    \caption{Illustration of the mapping between a schedule (or permutation) and a number in binary representation (top row) and the mapping for an example schedule with five operations (bottom row). The labels (i), (ii) and (iii) refer to the list at the beginning of \cref{sec:methods-A}.}
    \label{fig:map_constraint}
\end{figure*}

If we number the operations with labels $i\in\{1\cdots N_\mathrm{op}\}$, their position $s(i)$ in a schedule can be described by a permutation, $s(i)=\pi^{-1}(i)$. We denote the set of all schedules by $\mathcal{S}(\{1,\cdots,N_\mathrm{op}\})=\mathcal{S}_{N_\mathrm{op}}$. This set also contains schedules that do not obey \cref{eq:lexicographical} and is thus identical to the set of permutations, i.e. the symmetric group. There are $N_\mathrm{op}!$ unique schedules (permutations) in  $\mathcal{S}_{N_\mathrm{op}}$. To identify each schedule uniquely with a quantum state of a quantum processor, we need $\lceil\log_2(N_\mathrm{op}!)\rceil$ qubits and a map from their bitstring space $\{0,1\}^{\lceil\log_2(N_\mathrm{op}!)\rceil}$ to the set of possible schedules. We construct this map with the help of the inversion vectors of permutations and the factorial number system via the following three steps.
\begin{itemize}
    \item[(i)] Each schedule or permutation can be unambiguously identified by its inversion vector, an array of numbers that denote the number of inversions in the permutation for each position \cite{knuth_art}. 
    \item[(ii)] The elements of the $n$-element inversion vector of a permutation and the digits of an $n$-digit number in the factorial number system \cite{knuth_art} can both take the same number of possible values, one can label each inversion vector of length $n$ unambiguously with an $n$ digit number in the factorial number system.
    \item[(iii)] Each number in the factorial number system is a representation of an integer number, which can also be written as a bit-string in its binary representation.
\end{itemize}

Taking all these identifications together, we get a one-to-one mapping between the bit-strings represented on the qubits and the permutations which identify the schedules.
These steps are illustrated in \cref{fig:map_constraint}, where they are also labeled (i), (ii) and (iii) as in the enumeration above, and shown for a simple example.

We now explain the mapping in more detail, starting with a discussion of the required inversion vectors, which also permit a very elegant formulation of the job precedence constraint in \cref{eq:lexicographical}.

\subsubsection{Inversion vector}
To map a bit-string to a permutation, we make use of an intermediate step, the so called inversion vector $I(\pi)=(b_1,\cdots,b_{N_\mathrm{op}})$ with $b_j=\#\{k|k>j\wedge\pi^{-1}(k)<\pi^{-1}(j)\}$ \cite{knuth_art}. 

In the tuple representation of $\pi$, i.e. the array $(\dots, \pi(j), \pi(j+1),\dots)$, the value of $b_j$ denotes the number of elements to the left of $j$ that are larger than $j$. An example for the computation of the inversion vector is provided in \cref{tab:invvec}. Importantly the elements $b_j$ of $I(\pi)$ fulfill $b_j \leq N_{\text{op}}-j$ (We will make use of this property later). 
\begin{table}[h]
    \begin{tabular}{|r|c|l|c|}
    \hline
        $j$ & $\pi^{-1}(j)$ & $\pi(l) \text{ for } l < \pi^{-1}(j)$ & $b_j$ \\
        \hline\hline
         1 & 3 & $\pi(1)=3>1, \pi(2)=5>1 $ & 2 \\
         2 & 4 & $\pi(1)=3>2, \pi(2)=5>2, \pi(3)=1<2  $ & 2 \\
         3 & 1 &  & 0 \\
         4 & 5 & $\pi(2)=5>4  $ & 1 \\
         5 & 2 & $\pi(l)\leq 5\quad \forall l$ & 0\\
         \hline
    \end{tabular}

    \bigskip
    
    \begin{tabular}{|r|c|l|}
    \hline
        $j$ & $b_j$ & $\pi$ \\
        \hline\hline
         5 & $b_5=0$ & $\pi=(5)$\\
         4 & $b_4=1$ & $\pi=(5, 4)$\\
         3 & $b_3=0$ & $\pi=(3, 5, 4)$\\
         2 & $b_2=2$ & $\pi=(3, 5, 2, 4)$\\
         1 & $b_1=2$ & $\pi=(3, 5, 1, 2, 4)$\\
         \hline
    \end{tabular}
\caption{Example for a conversion between a permutation and its inversion vector. Top: Computation of inversion vector elements $b_j$ for the permutation $\pi=(3,5,1,2,4)$. 
Bottom: Computation of the permutation $\pi$ from its inversion vector $I(\pi)$.}
    \label{tab:invvec}
\end{table} 

Conversely, given the inversion vector $I(\pi)$ one can retrieve the permutation $\pi$ by iteratively determining the relative position of the numbers $j$ (which enumerate the operations $O_j$). Starting from $N_\mathrm{op}$,  for $j = N_\mathrm{op}, N_\mathrm{op}-1, \cdots, 1$, each $b_j$ determines the position of $j$ in the respective iteration, where positions are counted from the left starting from 0 (i.e. $b_j=0$ is the first and $b_j=1$ the second position etc.). This is also illustrated in \cref{tab:invvec} for the  same example.

Hence, each permutation can be unambiguously identified by its inversion vector and vice versa, see step (i) in \cref{fig:map_constraint}. For representing the problem on qubits, we further need a map from the set of bit-strings
to the inversion vectors. This can be obtained by using the factorial number system (FNS).

\subsubsection{Factorial number system}
The representation of a positive integer $x \in \mathbb{N}$ in the factorial number system (FNS) \cite{knuth_art} is composed of $n$ digits $c_k$ ($k=1, 2, \dots,n$)  with $n! > x$ and has the form
\begin{equation} \label{eq:FNS-number}
    x = \sum_{k=1}^n  c_k \, (k-1)! 
\end{equation}
where $ 0 \leq c_k < k$ and $k$ is also called the base for $c_k$. 
As for each number system, one can convert an integer number into the factorial number system by iteratively dividing by the base $k$, taking the remainder as the digit $c_k$ and continuing with the obtained integer quotient until this quotient becomes 0, c.f. \ref{sec:Invvec}. Here we want to use an FNS number to count all possible permutations of $N_\text{op}$ operations and thus choose $n=N_\text{op}$, see step (iii) in \cref{fig:map_constraint}.

The $k$-th digit of an FNS number can take on $k$ possible values, see \cref{eq:FNS-number}. The same holds for the inversion vector element $b_j$ with $j= N_{\text{op}}-(k-1)$, where $j,k = 1, \dots, N_{\text{op}}$. Therefore the elements $b_j$ are equivalent to the digits $c_k$, and we identify
\begin{equation}
    b_j = c_{N_{\text{op}}-(j-1)}.
\end{equation}
This identification
allows us to use  FNS numbers and inversion vectors interchangeably to label permutations unambiguously, corresponding to step (ii) in \cref{fig:map_constraint}.

Taking the above steps together, we thus have a complete map from $\{0,1\}^{\lceil\log_2(N_\mathrm{op}!)\rceil}$ to $S(\{1,\cdots,N_\mathrm{op}\})$. An example is provided in \cref{fig:map_constraint}.

We use this mapping to represent a schedule as a basis state of a register of qubits, e.g. $\ket{111101}$ for the example in \cref{fig:map_constraint}.
Note that up to this point we still consider all permutations/schedules, i.e. we also include schedules that do not obey \cref{eq:lexicographical}. We now present a modification of the above mapping that excludes invalid schedules.

\subsection{Resolving the job precedence constraint}\label{sec:job_con}
\begin{figure*}[t]
    \centering
    \scalebox{1}{
\begin{tikzpicture}[scale =1.5]
\shade[ball color=red!20!, fill opacity=0.75, text opacity= 1] (0,0) circle (1) node{$S^\mathcal{J}_{N_\mathrm{op}}$};
\shade[ball color=red!20!, fill opacity=0.75, text opacity= 1] (3,0) circle (1) node{$\{I(\pi)\}_{\pi\in S^\mathcal{J}_{N_\mathrm{op}}}$};
\shade[ball color=red!20!, fill opacity=0.75, text opacity= 1] (6,0) circle (1) node{JNS};
\shade[ball color=red!20!, fill opacity=0.75, text opacity= 1] (9,0) circle (1) node{$\{0,1\}^{\lceil\log_2(|S^\mathcal{J}_{N_\mathrm{op}}|)\rceil}$};
\draw[<->, very thick](1,0) -- (2,0) node [above,midway] {(i)};
\draw[<->, very thick](4,0) -- (5,0) node [above,midway] {(iv)};
\draw[<->, very thick](7,0) -- (8,0) node [above,midway] {(v)};
\end{tikzpicture}
}
    \caption{Illustration of the mapping between a number in binary representation and a valid schedule (or permutation) via the job number system (JNS), see \cref{eq:JNS}, that automatically respects the job precedence constraint.}
    \label{fig:map_noconstraint}
\end{figure*}
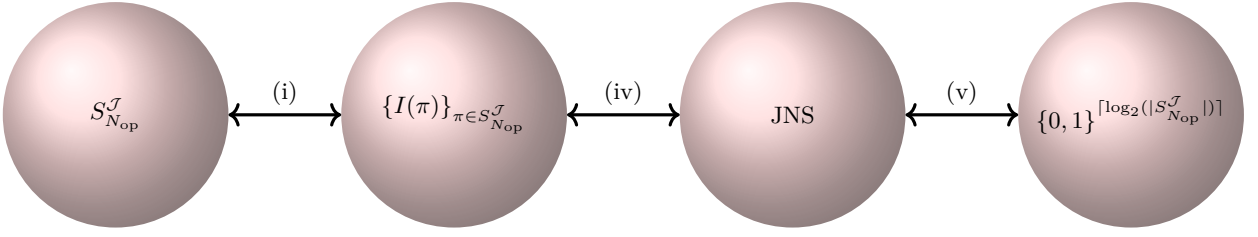
In the problems we consider, the order of the operations in one job is fixed and cannot be changed, c.f. \cref{eq:lexicographical}. A possible choice to avoid this would be to introduce a penalty term in the cost function for invalid schedules. Instead, we resolve the job precedence constraint of \cref{eq:lexicographical}, within the encoding. This saves resources since less qubits are needed on the quantum device. To resolve the constraint of \cref{eq:lexicographical}, we first need to translate it into inversion vector elements.

The job precedence constraint says that $\pi^{-1}(j) \leq \pi^{-1}(j+1)$ must hold for all $j $ and $ j+1$ in the same job. This translates to the constraint $b_{j+1}\geq b_j$ for the corresponding inversion vector elements, as we show in \cref{sec:precedence-proof}.

Since we want to use these constraints to reduce the code space, we need a new map from bit-strings to schedules that only counts the subset of allowed schedules
\begin{equation}
S^\mathcal{J}_{N_\mathrm{op}} \equiv S_{N_\mathrm{op}}\setminus \{\pi|\exists j, i:\pi^{-1}(j) > \pi^{-1}(j+1)\} 
\end{equation}
 for $j,j+1 \in J_i,\, i \in \{1,\dots,N_\mathrm{op}\}$, as depicted in \cref{fig:map_noconstraint}. Since there are $|J_i|!$ permutations of the operations within job $i$, but only one is allowed, the number of valid schedules reduces to
 \begin{equation}\label{eq:number_valid_schedules}
     |S^\mathcal{J}_{N_\mathrm{op}}|= \frac{N_\mathrm{op}!}{\prod_k |J_k|!}.
 \end{equation}
 
To construct a map to bitstrings, we first need to correctly assign bitstrings to inversion vector elements and then identify these with schedules as in step (i) above. We start by calculating the number of allowed inversion vector element combinations for each job, i.e. combinations for which $b_{j+1}\geq b_j$.
For a job $J_i=\{O_g,\dots,O_{g+l}\}$. c.f. \cref{eq:lexicographical} for $g=k_{i-1}+1$ and $g+l = k_i$, there are the associated inversion vector elements $b_g, \cdots, b_{g+l}$. Since each $b_j$ can take on $N_{\text{op}}-j+1$ possible values, there are $\prod_{j=g}^{g+l} (N_\mathrm{op} - j+1)$ possible values for the inversion vector elements corresponding to the job $J_i$. For a job of size $l+1$ the number of valid permutations is reduced by a factor $(l+1)!$ since there are $(l+1)!$ permutations of the operations within the job, of which only one is allowed. Since the constraint of a job $J_i$ only affects the digits in $b_g, \cdots, b_{g+l}$, there are 
\begin{equation} \label{eq:bj-values-per-job}
   N_\mathrm{digit}(\{b_g,\cdots,b_{g+l}\})= \frac{1}{(l+1)!} \prod_{j=g}^{g+l} (N_\mathrm{op} - j+1)
\end{equation}
possible values for these digits. One can retrieve the total number of allowed schedules, i.e. \cref{eq:number_valid_schedules}, by multiplying the vlaues of $N_\mathrm{digit}$ for all jobs,
\begin{equation}
\prod_i N_\mathrm{digit}(J_i) = \frac{N_\mathrm{op}!}{\prod_k |J_k|!}.
\end{equation}

Having determined the number of allowed schedules and inversion vector elements for a given set of jobs, we now turn to enumerate these allowed schedules with an appropriate number system, see step (v) in \cref{fig:map_noconstraint}, which replaces the former step (iii). A suitable number system is provided by the representation 
\begin{equation} \label{eq:JNS}
    x = \sum_{i=1}^{N_\mathrm{jobs}} r_i \prod_{m=1}^{i-1} N_{\text{digit}}(J_{N_\mathrm{jobs}-m+1}),
\end{equation}
where $0 \leq x \leq \prod_i N_\mathrm{digit}(J_i) - 1$ and the role of the factorials in \cref{eq:FNS-number} is now taken by the products $\prod_{m=1}^{i-1} N_{\text{digit}}(J_{N_\mathrm{jobs}-m+1})$.

We call the representation in \cref{eq:JNS} job number system (JNS). To obtain the values of the $r_i$, one follows a similar procedure as for the FNS and divides $x$ iteratively by the job sizes $N_{\text{digit}}(J_{N_\mathrm{jobs}-i+1})$, which play the role of the base here, taking the remainders of the division as the $r_i$ and continuing with the obtained quotient.

Each of the resulting $r_{N_\mathrm{jobs}-i+1}\in \{0, \dots, N_{\text{digit}}(J_i)-1\}$ in \cref{eq:JNS} thus enumerates the allowed values for the string $b_g, \cdots, b_{g+l}$ associated to job $J_i$. To use this relation to enumerate all schedules, we however still need to identify each value of $r_{N_\mathrm{jobs}-i+1}$ with a specific allowed string $b_g, \cdots, b_{g+l}$, see step (iv) in \cref{fig:map_noconstraint}, which replaces the former step (ii).

We do this by determining the values $b_j$ iteratively, starting with $b_g$. Its value $b_g^*$ needs to be chosen as the minimal $k$, such that the number of allowed strings $b_{g}, \cdots, b_{g+l}$ with $b_g\leq k$ is larger than $r_{N_\mathrm{jobs}-i+1}$. 

There are $N_{\text{digit}}(\{b_{g+1}, \dots , b_{g+l}\})$  strings with $b_g = 0$. For  $b_g = 1$ however $b_{j} \neq 0$ for $j = g+1, \dots, g+l$. Since each $b_j$ can take on $N_{\text{op}}-j+1$ possible values, the number of possible strings with $b_g = 1$ is identical to $N_{\text{digit}}(\{b_{g+2}, \dots , b_{g+l+1}\})$. This observation can be iterated across all values for $b_g$, where the number of possible strings for each fixed $b_g$ is $N_{\text{digit}}(\{b_{g+1 + b_g}, \dots , b_{g+l+ b_g}\})$.

After determining $b_g$ we iterate the same procedure over the remaining $b_j$. If all the already determined $b_g^*, b_{g+1}^*, \dots , b_{j-1}^*$ are zero, the number of possible strings for the remaining $b$s for $b_j=0$ is $N_{\text{digit}}(\{b_{j+1}, \dots , b_{g+l}\})$. In turn, if the $b_g^*, b_{g+1}^*, \dots , b_{j-1}^*$ are not all zero, but $b_j$ takes on its smallest possible value $b_j = m_j$, where $m_j = b^*_{j-1}$ for $j>g$ and $m_g=0$, the number of possible strings for the remaining $b$s reduces to $N_{\text{digit}}(\{b_{j+1+m_j}, \dots , b_{g+l+m_j}\})$.
If $b_j$ takes on a larger value than $m_j$, the number of possible strings for the remaining $b$s reduces further to $N_{\text{digit}}(\{b_{j+1+k'+m_j}, \dots , b_{g+l+k'+m_j}\})$, where $k' = b_j - b_{j-1}^*$.

Similar to $j=g$, the value $b_j^*$ is chosen with the minimal $k$ such that the number of possible strings with $b^*_{j-1}\leq b_j \leq b^*_{j-1} +k$,
\begin{align}
    & \tilde{N}(g,l,k,m_j,j) = \\
    & = \sum_{k'=0}^{k} N_\mathrm{digit}(\{b_{j+1+k'+m_j},\cdots,b_{g+l+k'+m_j}\}), \nonumber
\end{align}
is larger than
\begin{equation}
    r^\prime_{N_\mathrm{jobs}-i+1}= r_{N_\mathrm{jobs}-i+1}-\sum_{u=g}^{j-1} \tilde{N}(g,l,b_u^*-m_u-1,m_u,u), 
\end{equation}
where the number of already excluded strings $\sum_{u=g}^{j-1} \tilde{N}(g,l,b_u^*-m_u-1,m_u,u)$ needs to be subtracted.
Hence, the $b_j^*$ for $j \in \{g+1, \dots, g+l\}$ are determined by,
    \begin{equation}
             b_j^* = b_{j-1}^*+ \text{min}\left\{k \left|
    \tilde{N}(g,l,k,m_j,j) > r^\prime_{N_\mathrm{jobs}-i+1}  \right.\right\}
    \end{equation}

This procedure of determining the values  $b_j^*$, is also illustrated with an example of a job of size three in \cref{sec:conversion_example}. The number of queries, i.e. the number of calling \cref{eq:bj-values-per-job}, can be bounded for determining one $b_j$ by $N_\mathrm{op}-j+1$ s.t. that the total number of queries for converting all $r$s is bounded by $\mathcal{O}(N_\mathrm{op}^2)$.

After resolving the job precedence constraint, the number of required qubits for a JSP problem with $|\mu_i|=1\,\forall i$ is given by \cref{eq:nqubits_JSP}

\subsection{Operations with multiple possible machines}\label{sec:multm}
Up to this point we considered the reduced problem with $|\mu_i|=1\quad \forall i \in \{1,\cdots, N_\mathrm{op}\}$, i.e. every operation has exactly one machine that can execute it. We will now look at the more general case with $|\mu_i|\in \{1,\cdots, N_\text{ma}\}$. This implies that for every fixed schedule there are now $\prod_i |\mu_i|$ possibilities for machine configurations and the number of required qubits is given by \cref{eq:nqubits_FJSP}.

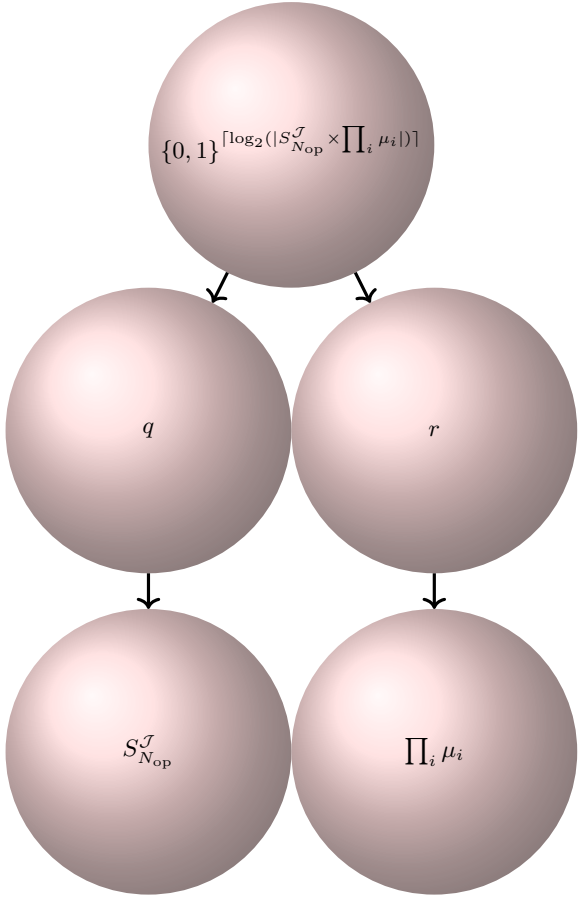
\begin{figure}[ht!]
    \Centering
    \begin{tikzpicture}[scale =1.9]
\shade[ball color=red!20!, fill opacity=0.75, text opacity= 1] (0,0) circle (1) node{$\{0,1\}^{\lceil\log_2(|S^\mathcal{J}_{N_\mathrm{op}}\times\prod_i \mu_i|)\rceil}$};
\shade[ball color=red!20!, fill opacity=0.75, text opacity= 1] (-1,-2) circle (1) node{$q$};
\shade[ball color=red!20!, fill opacity=0.75, text opacity= 1] (1,-2) circle (1) node{$r$};
\shade[ball color=red!20!, fill opacity=0.75, text opacity= 1] (-1,-4.25) circle (1) node{$S^\mathcal{J}_{N_\mathrm{op}}$};
\shade[ball color=red!20!, fill opacity=0.75, text opacity= 1] (1,-4.25) circle (1) node{$\prod_i \mu_i $};
\draw[->, very thick](-0.4472135955,-0.8944) to (-0.5527864045,-1.1056);
\draw[->, very thick](0.4472135955,-0.8944) to (0.5527864045,-1.1056);
\draw[->, very thick](-1,-3) to (-1,-3.25);
\draw[->, very thick](1,-3) to (1,-3.25);
\end{tikzpicture}
    \caption{Pictorial representation of converting a bitstring for a FJSP instance into a valid permutation and a machine configuration}
    \label{fig:multm_bubbles}
\end{figure}

Thus, we need a map from the binary system to a valid permutation and a valid machine configuration. To this end we convert the binary number to a decimal and divide by $\prod_i |\mu_i|$. We then use the integer quotient $q$ to determine the permutation of the operations as described in the previous section, see \cref{fig:map_noconstraint} and the remainder $r$ is used to get the machine configuration, depicted in \cref{fig:multm_bubbles}. Similar to the FNS and the JNS we can interpret the cardinality of the sets $\mu_i$ as a mixed base number system. So we divide the remainder $r$ by $|\mu_1|$ and the resulting remainder $r_1$ is the $r_1$-th element in $\mu_1$. We then divide the integer quotient by $|\mu_2|$ and iterate this procedure.

\clearpage
\section{Supplementary Information}
\subsection{Properties of the factorial number system} \label{sec:Invvec}

This appendix provides properties and examples for FNS representations.
The conversion of an integer from the decimal system to any mixed radix number system always follows the same scheme. Given the integer $x$ and the base $(b_n,\dots,b_1)$ of the first $n$ digits, the task is to determine the digits $(c_n,\dots,c_1)$ such that
\begin{equation}\label{eq:mixedraidx}
    x=\sum_{i=1}^n c_i \prod_{m=1}^{i-1} b_m.
\end{equation}
This can be achieved by iteratively dividing the integer with the base starting from $b_1$. Since \cref{eq:mixedraidx} can be written as 
\begin{equation}
    x=c_1+b_1\left[\sum_{i=2}^n c_i \prod_{m=2}^{i-1} b_m\right]
\end{equation}
we see that the right summand is a multiple of $b_1$ and since $c_1<b_1$ per definition, the remainder is equivalent to $c_1$. The next step consists of dividing the expression in the square brackets by $b_2$ in order to extract $c_2$. This is iterated until the divisor is larger than the dividend.

\Cref{tab:fns} shows properties of the FNS for up to eight digits.
\begin{table}[h!]
\begin{center}
\begin{tabular}{|c|| c| c| c| c| c| c| c| c|} 
 \hline
 base & 8& 7& 6& 5& 4& 3& 2& 1\\
 \hline
 place value & 7!& 6!& 5!& 4!& 3!& 2!& 1!& 0!  \\ 
 \hline
  place value in decimal & 5040&720& 120& 24& 6& 2& 1& 1  \\ 
 \hline
  highest digit & 7& 6& 5& 4& 3& 2& 1& 0  \\ 
 \hline
\end{tabular}
\caption{Factorial number system up to 8 digits.}
\label{tab:fns}
\end{center}
\end{table}

We further illustrate the conversion of a number in decimal system to its representation in the factorial number system for the example 61.
\begin{align*}
        61 \div 1 &= 61 &\text{ remainder } 0\\
        61 \div 2 &= 30 &\text{ remainder } 1\\
        30 \div 3 &= 10 &\text{ remainder } 0\\
        10\div 4 &= 2 &\text{ remainder } 2\\
        2 \div 5 &= 0 &\text{ remainder } 2\\
\end{align*}
Thus the eight digit representation of 61 in the FNS is given by $00022010$.

\subsection{Proof of the job precedence constraint in terms of inversion vector elements} \label{sec:precedence-proof}
The job precedence constraint states that $\pi^{-1}(j) \leq \pi^{-1}(j+1)$ must hold for all $j $ and $ j+1$ in the same job. This translates to the constraint $b_{j+1}\geq b_j$ for the corresponding inversion vector elements.
\begin{proof} We first show that $b_{j+1}\geq b_j$ for all cases, where $\pi^{-1}(j) < \pi^{-1}(j+1)$:
For all $i$ that generate inversions with $j$, i.e. $\forall i\in$  $\{i|i>j \wedge \pi^{-1}(i)<\pi^{-1}(j)\}$, it follows that
\begin{itemize}
    \item $i > j+1$ since  $i \neq j+1$ because $\pi^{-1}(j)<\pi^{-1}(j+1)$ per definition,
    \item  $\pi^{-1}(i)<\pi^{-1}(j+1)$ since $\pi^{-1}(i)<\pi^{-1}(j)$ and $\pi^{-1}(j)<\pi^{-1}(j+1)$.
\end{itemize}
Hence every $i$ that generates an inversion with $j$ also generates an inversion with $j+1$. Per definition of the inversion vector, this implies $b_{j+1}\geq b_j$.

It remains to show that $b_{j}> b_{j+1}$ whenever the constraint  $\pi^{-1}(j) < \pi^{-1}(j+1)$ is violated: 
For all $i$ that generate inversions with $j+1$, i.e. $\forall i\in$ $\{i|i>j+1 \wedge\pi^{-1}(i)<\pi^{-1}(j+1)\}$, it follows that
\begin{itemize}
    \item $i>j$ since $j+1 > j$
    \item $\pi^{-1}(i)<\pi^{-1}(j)$ since $\pi^{-1}(i)<\pi^{-1}(j+1)$ and $\pi^{-1}(j+1)<\pi^{-1}(j)$
\end{itemize}
This implies $b_{j}\geq b_{j+1}$. Furthermore, since $\pi^{-1}(j+1)<\pi^{-1}(j)$ per definition, which increases the inversion count of $j$ by 1, it follows that $b_{j}> b_{j+1}$ .
\end{proof}

\subsection{Conversion example from the job number system to inversion vector elements}\label{sec:conversion_example} 
In this appendix we will give an example for the conversion from the job number system (JNS), i.e. the representation given in \cref{eq:JNS}, to inversion vector elements. The example job we consider has three operations, which are enumerated in the lexicographical order, c.f. \cref{sec:problem_def}, $J_i =\{O_{N_\mathrm{op}-3}, O_{N_\mathrm{op}-2}, O_{N_\mathrm{op}-1}\}=\{O_{g}, O_{g+1}, O_{g+2}\}$, where the index $g=N_{\mathrm{op-}3}$ is used to be consistent with the notation for the general scheme used in \cref{sec:job_con}. In this example we choose the integer in the JNS corresponding to $J_i$ as $r_{N_\mathrm{jobs}-i+1}=8$.  We now start with the inversion vector element $b_g$ corresponding to operation $O_{N_\mathrm{op}-3}$. Using \cref{eq:bj-values-per-job} we see that there are 6 possible combinations of $b_{g+1}$ and $b_{g+2}$ for $b_g = 0$, c.f. \cref{tab:ex_cycle}. However, since $r_{N_\mathrm{jobs}-i+1}=8$, and we thus want the ninth combination (counting from 0), we conclude that $b_g > 0$. From \cref{eq:bj-values-per-job}, we see that there are 3 possibilities for $b_{g+1}$ and $b_{g+2}$ for $b_g=1$ and therefore, the total number of strings covered by $b_g=0$ and $b_g=1$ is 9. We thus conclude that $b_g=1$ because the ninth combination is not within the first 6 but within the first 9 possibilities. After having fixed $b_g$ the integer $r_{N_\mathrm{jobs}-i+1}$ has to be adjusted by the number of combinations that correspond to smaller integers and have been excluded, i.e. the number of combinations that start with $b_g=0$, therefore $r^\prime_{N_\mathrm{jobs}-i+1}=8-6=2$. We will again use \cref{eq:bj-values-per-job} to determine that there are 2 possibilities for $b_{g+1}=1$ and 1 possibility for $b_{g+1}=2$ and conclude that $b_{g+1}=2$. The last inversion vector element  $b_{g+2}$ is now also fixed. A pictorial version of this procedure is given in \cref{tab:ex_cycle}.
\begin{table}[!htbp]
\begin{center}
\begin{tabular}{|c| c c  c|} 
 \hline
 Possibilities & 4!&3!&2!\\ [0.5ex] 
 \hline\hline
 $b_g=0$ & 0&0&0 \\ 
  $N_\mathrm{digit}(\{b_{g+1}, b_{g+2}\})= \frac{4\cdot 3}{2!}$& 0&0&1 \\ 
  & 0&0&2 \\ 
  & 0&1&1 \\
  & 0&1&2 \\
  & 0&2&2 \\
 \hline
 $b_g=1 $ & 1&1&1 \\
  $N_\mathrm{digit}(\{b_{g+1+1}, b_{g+2+1}\})=\frac{3\cdot 2}{2!}$& 1&1&2 \\  
  & 1&2&2 \\  
 \hline
 $b_g=2$&&&\\
 $N_\mathrm{digit}(\{b_{g+1+2}, b_{g+2+2}\})=\frac{2\cdot 1}{2!} $\parbox[0pt][1.5em][c]{0cm}{}& 2&2&2 \\ 
  \hline
\end{tabular}\\[0.1cm]
\begin{gather*}
 \tilde{N}(g,2,0,0,g) = 6\\ \leq r_{N_\mathrm{jobs}-i+1} = 8 < \\\tilde{N}(g,2,1,0,g) = 9
 \end{gather*}
$\Downarrow$\\
$b_g^* = 1$\\
$\Big\downarrow$\\[0.1cm]
\begin{tabular}{|c|  c  c|} 
 \hline
 Possibilities &3!&2!\\ [0.5ex] 
 \hline\hline
$b_g^*=1, b_{g+1}=1$ &1&1 \\
$N_\mathrm{digit}(\{b_{g+2+0+1}\})=\frac{2}{1!}$\parbox[0pt][1.5em][c]{0cm}{} &1&2\\ 
\hline
 $b_g^*=1, b_{g+1}=2$&&\\
 $N_\mathrm{digit}(\{b_{g+2+1+1}\})=\frac{1}{1!}$\parbox[0pt][1.5em][c]{0cm}{}&2&2\\  
  \hline
\end{tabular}\\[0.1cm]
\begin{gather*}
 \tilde{N}(g,2,0,1,g+1) = 2\\ \leq r_{N_\mathrm{jobs}-i+1}- \tilde{N}(g,2,0,0,g)= 2 < \\\tilde{N}(g,2,1,1,g+1) = 3
 \end{gather*}
$\Downarrow$\\
$b_{g+1}^* = b_g^*+1=2$\\
$\Big\downarrow$\\[0.1cm]
\begin{tabular}{|c|   c|} 
 \hline
 Possibilities &2!\\ [0.5ex] 
 \hline\hline
$b_g^*=1, b_{g+1}^*=2, b_{g+2}=2$ & \\
$N_\mathrm{digit}(\{\})=1$  &2 \\ 
  \hline
\end{tabular}\\[0.1cm]
\begin{align*}
  r_{N_\mathrm{jobs}-i+1}&- \tilde{N}(g,2,0,0,g)\\
  &-\tilde{N}(g,2,0,1,g+1)=0\\& < N_\mathrm{digit}(\{\})=1
 \end{align*}
$\Downarrow$\\
$b_{g+2}^* = b_{g+1}^*+0=2$\\
\caption{Pictorial representation for an example conversion from the JNS to inversion vector elements. The job $J_i =\{O_{N_\mathrm{op}-3}, O_{N_\mathrm{op}-2}, O_{N_\mathrm{op}-1}\}=\{O_{g}, O_{g+1}, O_{g+2}\}$ consists of three elements and its corresponding integer in the JNS is chosen as $r_{N_\mathrm{jobs}-i+1}=8$. 
The inversion vector elements are determined iteratively from top to bottom. The factorials at the top of the tables are the place values of the corresponding FNS digits and below are the possible digit combinations respecting the job constraint.
}
\label{tab:ex_cycle}
\end{center}
\end{table}

\subsection{Qubit scaling for various problem versions} \label{sec:appendic_scaling}
\subsubsection{Time indexed version}
In the time indexed formulation of the JSP \cite{Venturelli} one assigns a set of binary variables for each operation that indicate its starting time.
\begin{equation}\label{eq:timeindexedJSP}
    x_{i,t} = \begin{cases}
  1  & \text{operation $O_i$ starts at time $t$}\\
  0 & \text{else}
\end{cases}
\end{equation}
This results in $N_\mathrm{op}T$ qubits, where $T$ is a bound for $t\in\{0,\dots,T-1\}$ that represents a maximum time that one allows for the makespan. $T$ can be bounded from above by the sum over all execution times,
\begin{equation}
    T_\mathrm{max}=\sum_{i=1}^{N_\mathrm{op}} p_i ,
\end{equation}
and lower bounded by the minimum time required for an individual job,
\begin{equation}
    T_\mathrm{min}=\max_i \sum_{j=k_{i-1}+1}^{k_i} p_j .
\end{equation}
One could potentially increase the lower bound by looking at minimum time a machine is in action, however for simplicity, especially in FJSP case, this is disregarded. The number of qubits can be reduced by a simple variable pruning. An operation $O_l$ can never start before its predecessors have finished. This is called \textit{head} of the operation and is given for operation $O_l$ in job $J_i$ by.
\begin{equation}
    r_l := \sum_{k_{i-1}+1\leq j < l}p_j
\end{equation}
Analog to this, an operation needs to leave enough room for its successors to finish before the deadline $T$. This is called the \textit{tail} of the operation.
\begin{equation}
    q_l := \sum_{l< j \leq k_i}p_j
\end{equation}
Therefore, the number of possible starting times for operation $O_l$ is reduced to $t_l\in\{r_l,T-q_l-p_l\}$. The number of variables pruned by considering this within one job is given by:
\begin{align}
    \sum_{l}\left[r_l+q_l+p_l-1\right]&=\sum_{l}\left[p_l(|J_i|-1)+p_l-1\right]\nonumber\\
    &=\sum_{l}\left[p_l|J_i|-1\right]
\end{align}
This results in a total number of qubits for the time indexed version of
\begin{equation}\label{eq:ti_jsp}
    N_\mathrm{qubits}^\mathrm{ind}=N_\mathrm{op}T-\sum_i\sum_{l}(p_l|J_i|-1)
\end{equation}
with $i\in\{1,N_\mathrm{jobs}\}$ and $l\in\{k_{i-1}+1,\dots,k_i\}$.
For the time indexed formulation of the FJSP \autoref{eq:timeindexedJSP} is expanded to\cite{timeindex5}:
\begin{equation}\label{eq:timeindexedFJSP}
    x_{i,t,m} = \begin{cases}
  1  & \text{$O_i$ starts at time $t$ on machine $m$}\\
  0 & \text{else}
\end{cases}
\end{equation}
Then the number of qubits required become:
\begin{equation}
    N_\mathrm{qubits}^\mathrm{ind}=\sum_{j=1}^{N_\mathrm{op}} (T-r_j-q_j-p_j+1) |\mu_j|
\end{equation}
\subsubsection{Examples for problem instances}
\noindent\textbf{{Even scaling ($\mathbf{N\times N \times N}$):}}\\
The first example we will consider is the $N\times N \times N$ version of the JSP, in which there are $N$ jobs and machines and each job consists of $N$ operations, i.e. $N_\mathrm{op}=N^2$. Here \autoref{eq:nqubits_JSP} becomes:
\begin{equation}
    N_\mathrm{qubits}=\lceil \log_2(\frac{N^2!}{N!^N})\rceil\approx N^2 \log_2(N)
\end{equation}
The number of qubits for the indexed version, c.f. \cref{eq:ti_jsp},
can be upper bounded by $N_\mathrm{qubits}^\mathrm{ind} \leq <p>[N^4-N^3]+N^2$
and lower bounded by   $N_\mathrm{qubits}^\mathrm{ind} \geq [\max_i<p>_i-<p>]N^3+N^2 $, here $<p>_i$ denotes the average of execution times of job $J_i$.
The expected value is $<p>(A-1) N^3 + N^2$
with $<C_\mathrm{max}>=T\approx A N <p>$ and $A > 1$ \cite{Venturelli}.
The lower bound is the trivial case, especially for $\max_i <p>_i = <p>$ where every starting time is already determined before running the actual algorithm. It is important to note that the time indexed formalism depends on the spread of execution times $p$ which can introduce a factor 100 in required qubit number for realistic problem instances, that is quite relevant for NISQ devices.
If we now expand this example to a totally flexible job shop problem, i.e. $|\mu_i|=N$ for all $i$, the leading order of the qubit count for our approach is increased by a factor of 2, while the leading order of the time indexed version is multiplied by $N$.

\noindent\textbf{Growing Job sizes:} 
For the second example we consider instances where the number of jobs $N_\mathrm{jobs} = c$ is constant and the job sizes $|J_i|=N \: \, \forall i \in \{1,\dots,c\}$ grow evenly with system size.
For our approach this results in
\begin{align}
    N_\mathrm{qubits}&=\lceil \log_2(\frac{(cN)!}{N!^c})\rceil=\lceil\log_2((cN)!)-c \log_2(N!)\rceil\nonumber\\
    &=\lceil c N\log_2(c)+ \mathcal{O}(\log(N))\rceil
\end{align}
For the time index version, assuming $<C_\mathrm{max}>=T\approx A N <p>$ with $A > 1$, one gets:
\begin{equation}
    N_\mathrm{qubits}^\mathrm{ind}= (A-1)c N^2 <p>+ c N
\end{equation}
Expanding this example to a totally flexible job shop problem with a constant number of machines $c_\mathrm{ma}$ changes the prefactors of $c$ to $c+c_\mathrm{ma}$ for our approach and to $c\cdot c_\mathrm{ma}$ for the time indexed version. If the number of machines grows with system size, i.e. $N_\mathrm{ma}=N$, then the leading orders become $c N \log_2(N)$ and $(A-1)cN^3$, respectively.

\subsection{F-VQE}\label{sec:fvqe}
Since the FJSP is a combinatorial optimization problem, and we use a variational quantum algorithm for solving it, we need an optimization strategy for our algorithm. Due to its recent success in this field \cite{filtering, filtering_case_study} and its ability to be used for black box Hamiltonians, we chose to use
the Filtering Variational Quantum Eigensolver (F-VQE), see \cite{filtering}. This algorithm optimizes a parameterized quantum circuit to approximate the action of a filter operator.
\subsubsection{Filtering operators}
Given a Hamiltonian $\mathcal{H}$, note that in our case we will be using a black-box Hamiltonian as specified in \cref{sec:cost}, a filtering operator $F \equiv f(\mathcal{H};\tau)$ is defined by the real-valued function $f(E;\tau)$, where $E$ is the energy and $\tau > 0$. The square of this function is required to be strictly decreasing for the energy range $E \in [E_\mathrm{min}, E_\mathrm{max}]$.

Given a state $\ket{\psi}$ the probability to measure an eigenstate $\ket{\lambda_x}$ is $P_\psi(\lambda_x)=|\braket{\psi|\lambda_x}|^2$. Applying the filtering operator yields the state $\ket{F\psi}=F\ket{\psi} \big/ \sqrt{\left<F^2\right>_\psi}$, which now has a new probability distribution
\begin{equation}
    P_{F\psi}(\lambda_x)=\frac{f^2(E_x;\tau)}{\left<F^2\right>_\psi}P_\psi(\lambda_x)
\end{equation}
with an increased probability to measure eigenstates with $f^2(E_x;\tau)>\left<F^2\right>_\psi$, and hence increased amplitude of energetically low lying eigenstates. If the state $\psi$ has a finite overlap with the ground state, applying the filtering operator often enough results in a projection onto the ground state. We will here be using $f(\mathcal{H};\tau)=\mathcal{H}^{-\tau}$ as our filtering operator.

\subsubsection{Core of the algorithm}
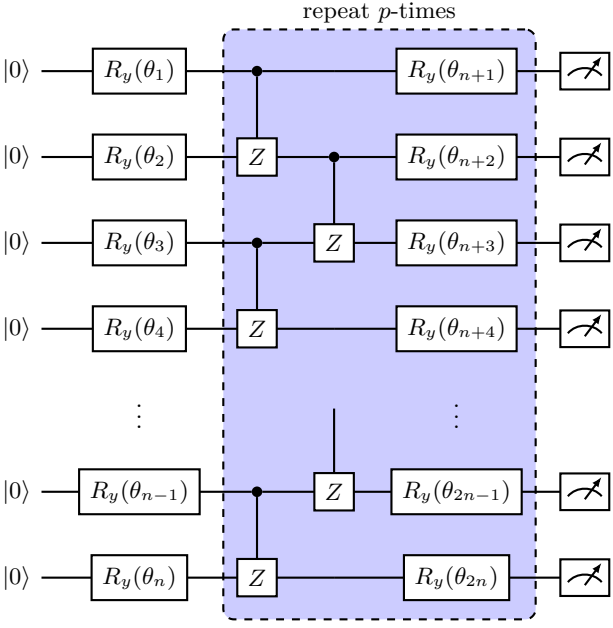
\begin{figure}[h]
\begin{equation*}
\begin{quantikz}
    \lstick{$\ket{0}$}&\gate{R_y(\theta_1)}&\ctrl{1}\gategroup[wires=7,steps=3,style={dashed,rounded corners,fill=blue!20, inner xsep=2pt},background]{{ repeat $p$-times}}&\qw&\gate{R_y(\theta_{n+1})}&\meter{}&\\
    \lstick{$\ket{0}$}&\gate{R_y(\theta_2)}&\gate{Z}&\ctrl{1}&\gate{R_y(\theta_{n+2})}&\meter{}&\\
    \lstick{$\ket{0}$}&\gate{R_y(\theta_3)}&\ctrl{1}&\gate{Z}&\gate{R_y(\theta_{n+3})}&\meter{}&\\
    \lstick{$\ket{0}$}&\gate{R_y(\theta_4)}&\gate{Z}&\qw&\gate{R_y(\theta_{n+4})}&\meter{}&\\
    &\vdots&&&\vdots&&\\
    \lstick{$\ket{0}$}&\gate{R_y(\theta_{n-1})}&\ctrl{1}&\gate{Z}\vqw{-1}&\gate{R_y(\theta_{2n-1})}&\meter{}&\\
    \lstick{$\ket{0}$}&\gate{R_y(\theta_{n})}&\gate{Z}&\qw&\gate{R_y(\theta_{2n})}&\meter{}&
\end{quantikz}
\end{equation*}
\caption{Parameterized quantum circuit used during the F-VQE; The gates in the blue box can be repeated $p$ times, where we used $p=1$ in our simulations.}
\label{fig:vqe_circ}
\end{figure}

The iterative algorithm starts at iteration step $t = 0$ and prepares the initial state $\ket{\psi_0}=\ket{+}^{\otimes n}$, where $n$ is the number of qubits. This state has finite overlap with the ground state. We want to approximate the repeated action of a filtering operator on it. At each optimization step $t \geq 1$ we approximate the action of a filtering operator $\ket{F_t\psi_{t-1}}$ resulting in state $\ket{\psi_t}$. Note that the filtering operator can change in each optimization step $t$. The algorithm terminates after a prespecified number of iteration steps and uses, a parameterized quantum circuit ansatz $\ket{\psi(\bm{\theta})}$, depending on $m$ variational parameters $\bm{\theta}= (\theta_1,\cdots,\theta_m)$. Our choice for the ansatz or gate sequence is shown in \cref{fig:vqe_circ}.

In a given optimization step $t$ one minimizes the Euclidean distance
\begin{equation}\label{eq:cost_t}
    \mathcal{C}_t(\bm{\theta)}=\frac{1}{2} ||\ket{\psi(\bm{\theta})}-\ket{F_t \psi_{t-1}}||^2=1-\frac{\mathrm{Re}\bra{\psi_{t-1}}F_t\ket{\psi(\bm{\theta})}}{\sqrt{\left< F_t^2\right>_{\psi_{t-1}}}},
\end{equation}
where the parameters found in this optimization process define $\ket{\psi_t}\equiv\ket{\psi(\bm{\theta}_t)}$. To optimize the cost function in \cref{eq:cost_t}, the F-VQE algorithm uses a gradient-based procedure, where the partial derivative of the cost function evaluated at the current vector of parameters $\bm{\theta}_{t-1}$ is given by:

\begin{equation}
    \left.\frac{\partial  \mathcal{C}_t(\bm{\theta)}}{\partial \theta_j}\right|_{\bm{\theta}_{t-1}}=-\frac{\left< F_t\right>_{\psi^{j+}_{t-1}}-\left< F_t\right>_{\psi^{j-}_{t-1}}}{4\sqrt{\left< F_t^2\right>_{\psi_{t-1}}}}
\end{equation}
The three states $\ket{\psi_{t-1}}$ and $\ket{\psi^{j\pm}_{t-1}}\equiv\ket{\psi(\bm{\theta_{t-1}}\pm \frac{\pi}{2}\bm{e}_j)}$ are generated by the circuit in \cref{fig:vqe_circ} with different parameters. The optimization at fixed $t$ is then completed by a single gradient-descent update: 
\begin{equation}\label{eq:update}
    \bm{\theta}_t=\bm{\theta}_{t-1}-\eta\sum_{j=1}^m \left.\frac{\partial  \mathcal{C}_t(\bm{\theta)}}{\partial \theta_j}\right|_{\bm{\theta}_{t-1}}\mathbf{e}_j 
\end{equation}
Here $\eta > 0$ is the learning rate, which is set in each iteration step to inverse of the cost function's Hessian diagonal. For each optimization step $t$ the update in \cref{eq:update} requires the preparation of $2m +1$ circuits.
Sampling eigenstates $\ket{\lambda_x}$ for $M_x$ times leads to the approximation of the expectation values as:
\begin{equation}
    \left< F_t\right>_{\psi}\approx\frac 1 M \sum_x M_x f(E_x;\tau)
\end{equation}
\subsubsection{Adapting the step size}
To ensure that the gradient norm $g(\tau)=||\nabla \mathcal{C}_t(\tau)|_{\bm{\theta}_{t-1}}||$ does not vanish, the parameter $\tau$ is adapted at every optimization process. At the beginning one sets a control value $g_c>0$. Since $F(\mathcal{H}; 0) = 1$ we know that $g(0) = 0$. We start at $\tau =0 $ and increases the value of $\tau$ until $g(\tau)> g_c$ or $g$ converges at a constant. We then choose the value for $\tau$ that results in a value of $g$ closest to the control value.

\begin{figure*}
 \Centering
\includegraphics[width=\textwidth]
{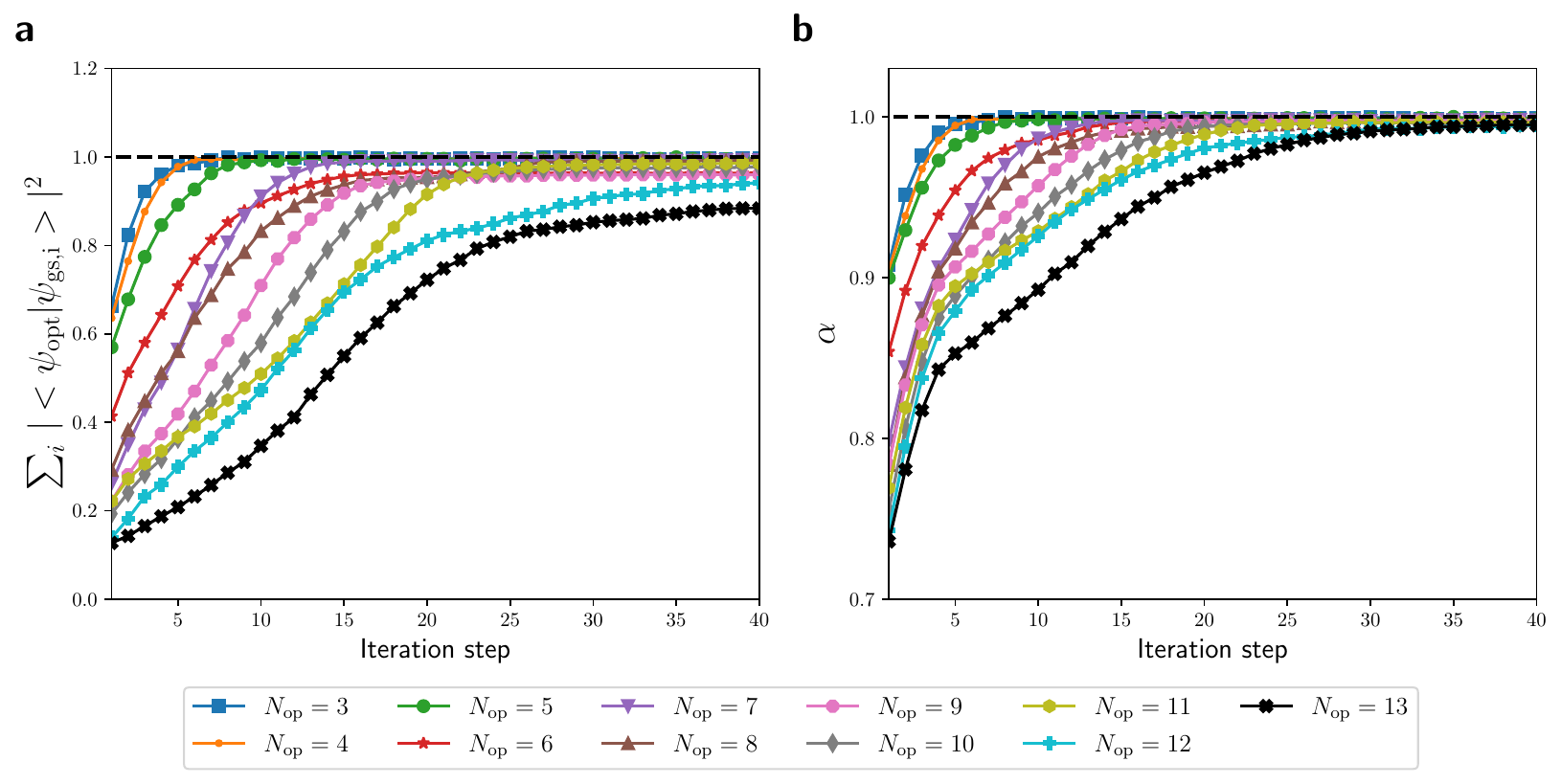}%
\caption{(\textbf{a}) Probability to measure the ground state, see \cref{eq:ground_state} and (\textbf{b}) approximation ratio $\alpha$, see \cref{eq:appr-ratio}, for JSPs with varying number of operations $N_\mathrm{op}$. We consider $N_\mathrm{meas}$ measurements for each set of variational parameters, where for $N_\mathrm{op}<6$ we use $N_\mathrm{meas}=100$, for  $6\leq N_\mathrm{op}<9$ we use $N_\mathrm{meas}=500$ and for $N_\mathrm{op}\geq 9$ we use $N_\mathrm{meas}=1000$. The legend applies to both plots. The data points are averages over 30 random JSP instances, c.f. \cref{sec:rand_inst}.}
\label{fig:permj_all}   
\end{figure*}

\subsection{Generating random problem instances}\label{sec:rand_inst}

To generate random JSP instances, we picked integer execution times $p_i \in \{1,\cdots, N_\mathrm{op}\}$ with equal probability for a fixed number of operations $O_i$ (The execution times could also be chosen to be floats.).
We took the number of operations in the first job $J_1$ randomly from a Gaussian distribution $\lfloor\mathcal{N}(\mu,\sigma)\rfloor$ with expectation value $\mu = \sqrt{N_\mathrm{op}}+\frac 1 2$ and standard deviation $\sigma = (\sqrt{N_\mathrm{op}}-\frac{1}{2})/3$, to ensure that $|J_i|$ is likely to be larger or equal to 1. For the case that it was smaller than 1 (or larger than $N_\mathrm{op}$), we rounded it to 1 (or $N_\mathrm{op}$).
We iterated the choice of jobs until all operations were allocated to jobs.
We also randomly chose the number of machines from $\lfloor \mathcal{N}(\sqrt{N_\mathrm{op}}+\frac{1}{2}, (\sqrt{N_\mathrm{op}}-\frac{1}{2})/3)\rfloor$ and randomly associated operations to the machines.

For the variational algorithm we took a fixed shot count depending on system size,  capped the iteration number to 30, chose $g_c=0.1$ and dynamically updated $\eta$, see also\cite{filtering}. We then average over 30 of such random instances.

For simulating FJSP instances we used the same procedure to generate jobs and execution times as for JSPs. This also means that $p_{i, M}= p_{i, M^\prime}\, \forall M, M^\prime \in \mu_i$, which is however not necessary for the algorithm. After determining the number of available machines $N_\text{ma}$, we chose the number of machines for each operation $|\mu_i|$ from $\lfloor \mathcal{N} (\sqrt{N_\text{ma}}+\frac 1 2, (\sqrt{N_\text{ma}}-\frac{1}{2})/3)\rfloor$ and randomly picked the elements of $\mu_i$ from the available machines.

\nocite{*}

\bibliography{bibliography}

\end{document}